# Comment to the article "Temperature dependence of ultracold neutron loss rates" E.Korobkina et al., PRB 70,035409.

A. Serebrov

*Petersburg Nuclear Physics Institute, Russian Academy of Sciences,
Gatchina, Leningrad District, 188300 Russia*

## Abstract

In work [1] ("Temperature dependence of ultracold neutron loss rates" E.Korobkina et al., PRB 70,035409) results of measurement of temperature dependence of losses of ultracold neutrons (UCN) in a range from 4 K to 300 K at UCN storage in a copper trap are presented. At interpretation of experimental data it was artificially considered only the difference of losses rate $1/\tau(T) = 1/\tau_{exp}(T) - 1/\tau_{exp}(10K)$, ($\tau$ - UCN storage time in the trap). It has been accepted for temperature dependence of losses which changes from 0 to $3.3 \cdot 10^{-4}$ per one collision of UCN with a trap surface. However, the analysis of raw experimental data shows that in a trap at temperature 10 K there is losses $1.0 \cdot 10^{-3}$ per collision. They are considerably (in 3 times) more than discussed temperature dependence and almost 10 times more than losses due to capture cross section on copper. It is the most probable that these losses are connected with leakage of UCN through a slit of trap shutter. Change of the size of a slits on 25 % at change of temperature from 300K to 10K can quite explain discussed temperature dependence. Certainly, hydrogen is present on a surface, but as it was shown at work [2] temperature dependence in 3 times lower, even on the undegased surfaces than on the degassed and deuterated surface in work [1]. At last, deuteration of a trap surface allows almost completely to suppress temperature dependence connected with presence of hydrogen [2]. The conclusion of work [1] is that hydrogen is localized on a surface in the form of a film, instead of distribution in the surface substance. This conclusion contradicts results of measurement of energy dependence of UCN losses in work [2]. More detailed analysis of work [1] and works [2] is presented below.

In Fig. 1 the scheme of experimental installation from work [1] is shown. The citation of the description of experimental installation is below. "We do not use separate vacuum but the whole apparatus was constructed to ultrahigh-vacuum standards, i.e., only metal seals were used (CF flanges with Cu gaskets and wire sealing with annealed Al, In, and Au); no plastic parts, only metal and ceramic; dry pump system (turbo-molecular pump and scroll forepump, Varian). The cryostat has five free outlets with CF-100 flanges around the main vacuum housing. One was used to connect a neutron guide through 100 mm Al foil. The turbomolecular pump was also mounted directly on another CF-100 outlet. In addition we have another cryostat working as a cryopump that is connected through an outlet 20 cm long, 25 cm in diameter to the main housing. Both cryostats are of the same construction (Oxford Instruments, UHV modification), but the UCN cryostat is attached to the UCN bottle and the cryopump cryostat is connected to large-area Cu baffles connected to both liquid nitrogen (LN) and liquid helium (LHe) baths. Thus we have very high-efficiency pumping system. Both heating and cooling were performed by direct contact of the storage bottle with a central part of the cryostat, which contained either a heating resistor or liquid nitrogen or liquid helium. Thus during heating the storage bottle was the hottest part of the apparatus whereas during cooling and at room temperature the coolest part was the cryopump filled with LN and LHe."

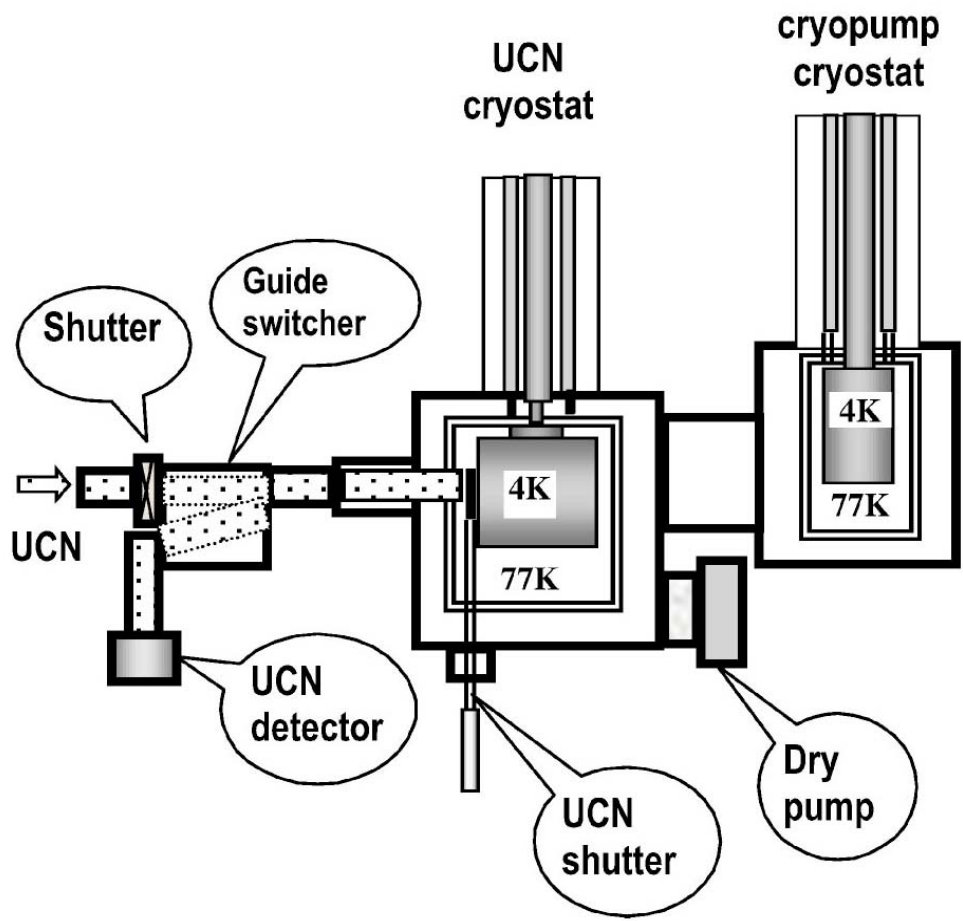

Fig. 1. Experimental layout from work [1].

In Fig. 2 temperature dependence of UCN losses rate from work [1] is shown, however the dependence is presented in full scale from zero and on the right axis the factor of UCN losses per collision is specified in addition. The factor of losses is obtained by dividing of loss rate into effective frequency of collision 65 Hz from work [1]. The factor of losses due to UCN capture cross section by Cu are estimated by value $1.7 \cdot 10^{-4}$, it is presented by line 1 in Fig. 2. For comparison the results from the work [2] is presented also.

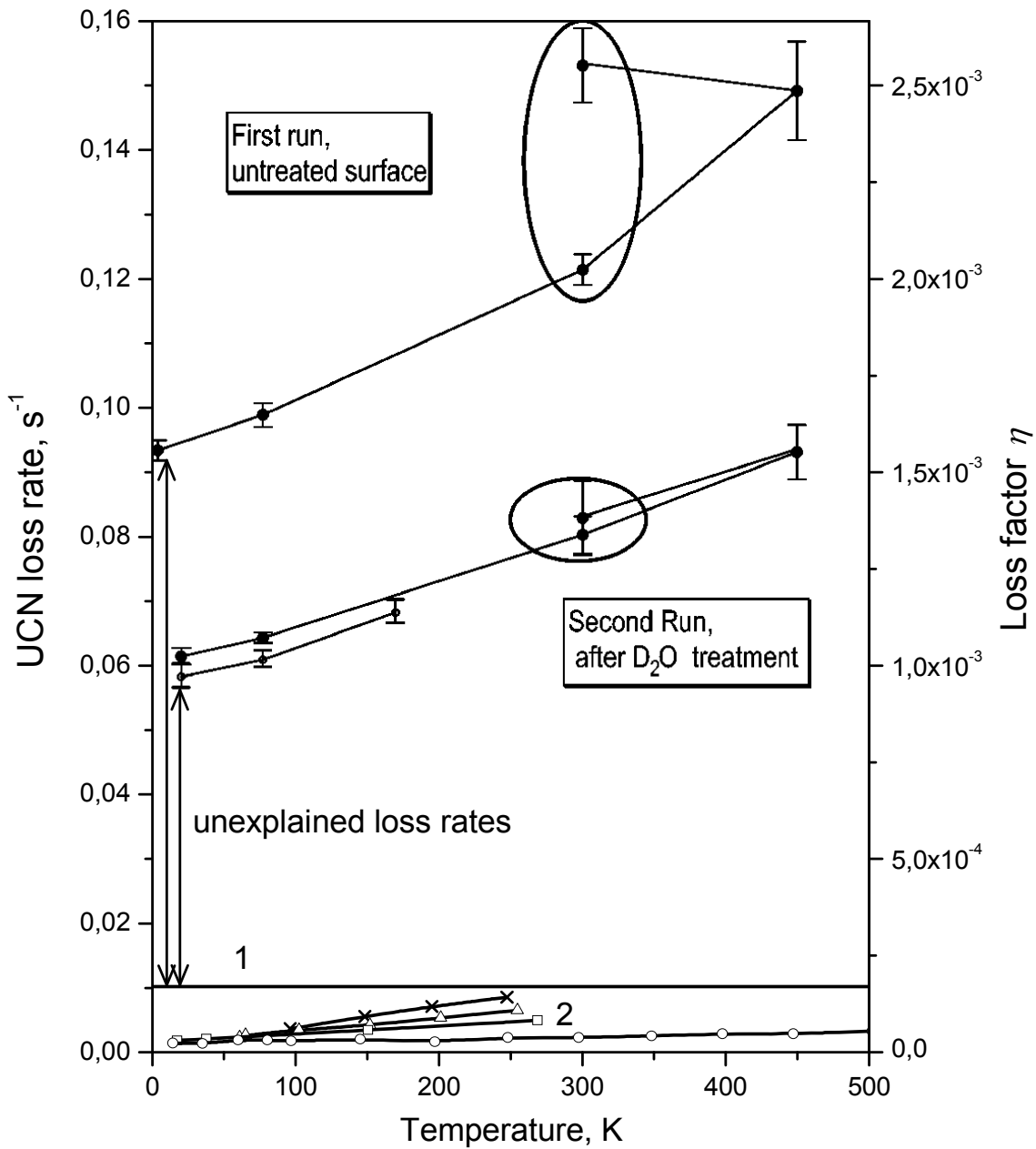

Fig. 2. Raw data for two different surface treatments prior to the measurement with UCN's: the upper curve, run 1, the surface was washed in the ultrasonic bath with distilled water; the lower two curves run 2, the surface was heated in $D_2O$ vapor in the vacuum oven. The ovals show the comparison of 300 K data before and after heating of the storage bottle inside the UHV cryostat.
1: UCN loss rate and corresponding loss factor due to capture cross section of Cu; 2: temperature dependences of loss factor on the beryllium surface from work [2].

The difference between experimental points at temperature 10K and the factor of losses due to capture cross section is about $0.8 \cdot 10^{-3}$. This difference is not discussed in any way in work [1]. After deuteration of trap surfaces the parallel shift of temperature dependence is observed though necessary expect change of an inclination of a curve because of hydrogen removal. This unclear effect also is not discussed in work [1] though losses at temperature 10K were decreased by factor of 1.7.

The most probable reason of unexplained losses is UCN leakage through slit of a shutter of the trap. The UCN leakage through other slits is possible also if the trap consists of several elements. In that case in experiment [1] temperature dependence of the size of slit together with temperature dependence of UCN losses were rather possibly studied. For example, reduction of the size of slit by 25 % at cooling of a trap from 300K to 10K can completely explain the observed temperature dependence. Thus, possibility of studying of temperature dependence of UCN losses on trap surface by means of installation [1] is represented to be extremely doubtful. Nevertheless, authors of work [1] try to make a choice between models of hydrogen sorption. The quantity of hydrogen (H) on a surface has been measured in independent experiment (elastic recoil detection analysis, ERDA). It gave $5.8 \cdot 10^{16}$ H/cm$^2$. However, ERDA does not give the information on hydrogen distribution on depth. In work [1] two models were considered: the specified quantity of hydrogen is distributed in substance with concentration of 8 % or it is on a surface in the form of a film in the thickness ~100A.

The Fig. 3 of work [1] shows that the model of the distributed hydrogen in substance cannot explain experimental dependence. The Fig. 4 of work [1] shows the consent with film model. The Fig. 4 misinforms the reader, showing that the probability of losses tend to zero with tendency of temperature to zero. Actually this result is obtained by simple subtraction of unexplained losses. If the effect of slit leakage would be measured and considered, then it is possible to discuss whether there are still temperature-independent losses except capture cross section. Unfortunately from [1] it is impossible to say anything about what part of these losses is temperature-independent, and what part is connected with leakage through a shutter. If to follow model of the distributed hydrogen where temperature dependence gives very small contribution then the basic part of temperature dependence in work [1] is connected with temperature dependence of the size of slit. However, the conclusion of work [1] becomes in favour of a film. Unfortunately it is impossible to accept the conclusions of given work even if to assume that after warming up the shutter slit was stable in size and did not vary at temperature change, and the found out temperature dependence concerns hydrogen on a trap surface. The matter is that the obtained results and conclusions contradict results of work [2], as however, and other works (see [3]). In this connection the short description of installation and results from work [2] is given below.

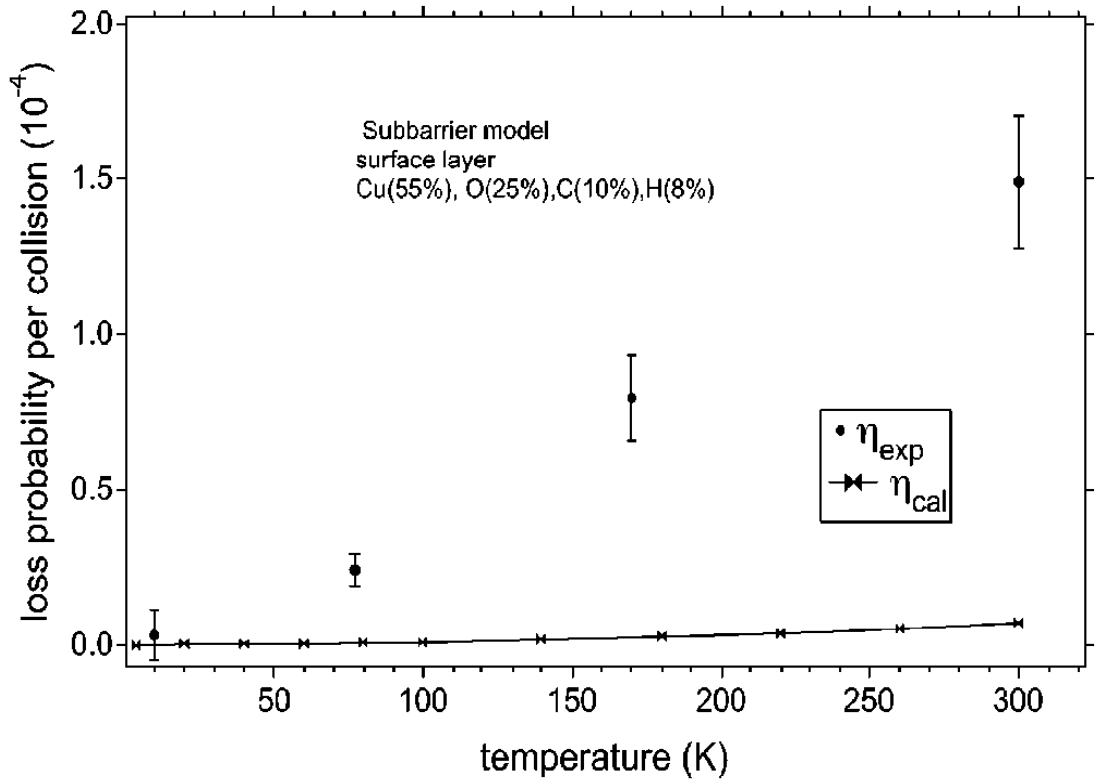

Fig. 3. Upscattering probabilities per collision for the subbarrier model: $\eta_{\mathrm{expt}}(T)$ is derived from experiment and $\eta_{\mathrm{calc}}(T)$ is calculated using $\sigma_{ie}(T)$ for hydrogen bound to Cu.

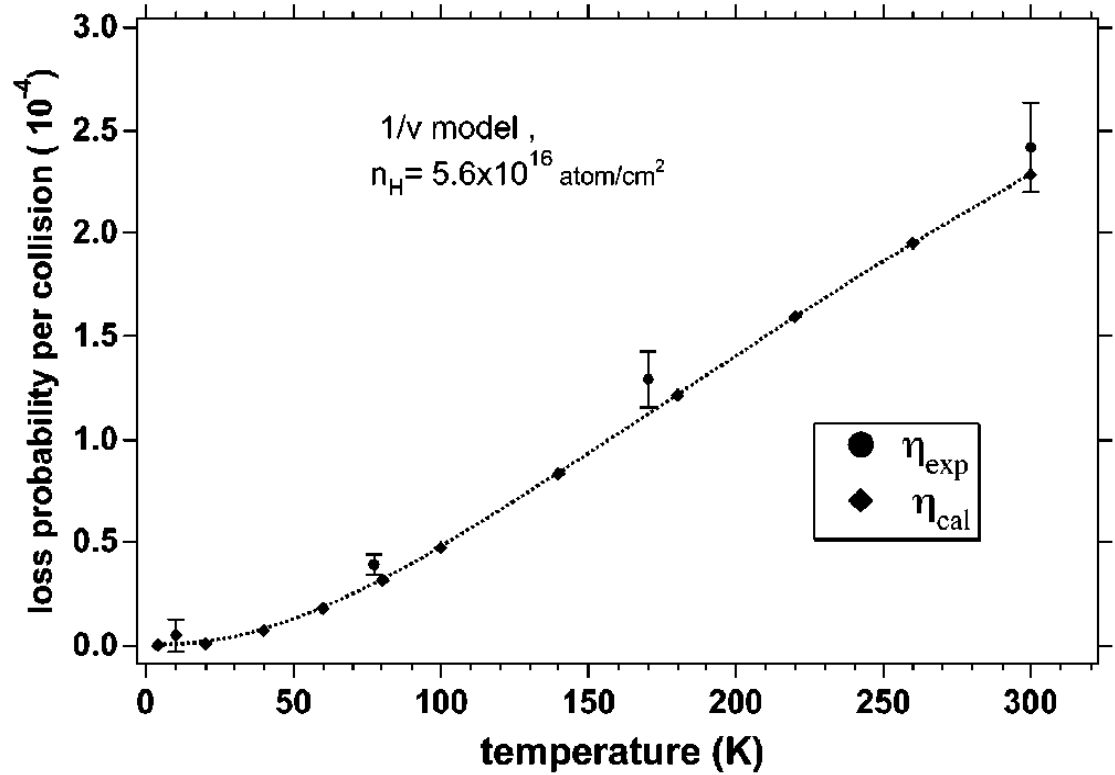

Fig. 4. Loss probability per collision calculated with $\sigma_{ie}(T)$ of ice and derived from experiment for the film model.

The installation scheme is presented in Fig.5. The experiment comprises a gravitational trap for UCN, which can also serve as a differential gravitational spectrometer. Hence, a distinctive feature of this experimental setup is its possibility of measuring the UCN energy spectrum after the neutron has been stored in the trap. The UCN storage trap (8) is placed inside the vacuum volume of the cryostat (9). The trap has a window and can be rotated around the horizontal axis in such a way that the UCN are held by the gravitational field in the trap when the window is in its uppermost position. UCN enter the trap through the neutron guide (1), after passing the inlet valve (2) and the selector valve (3). Filling of the trap by the ultra cold neutron gas is done with the trap window in the "down" position. After filling, the trap is rotated 180° so that the window is in the "up" position. The vacuum system comprises two separate vacuum volumes: the "high-vacuum" volume and the "isolating" volume. The pressure in the high-vacuum volume of the cryostat is $5 \cdot 10^{-6}$ mbar. The trap is cooled through the heat exchange between the trap and the cryostat's reservoir. To improve the heat exchange, gaseous helium was blown through the vacuum volume of the cryostat and later removed before measuring the storage time. The position (height) of the trap window, with respect to the bottom of the trap, determines the maximum energy of the UCN that can be stored in the trap. The different values of the window's height correspond to the different values of the cutoff energy in the UCN spectrum; this rotating trap is thus a gravitational spectrometer. The spectral dependence of the neutron storage time can be measured by a series of measurements whereby one varies the trap window height. The trap was kept in each position for 100–150 s to register the UCN in that energy range. In according to this procedure we measured UCN storage time as function of energy. Two UCN traps with different dimensions were employed. The first trap was quasi-spherical, consisting of a horizontal cylinder of 26 cm in length and 84 cm in diameter that was "crowned" by two 22-cm-high truncated cones with a smaller diameter of 42 cm. The second trap was cylindrical, with the length of 14 cm and 76 cm in diameter. The frequency of neutron collisions with the walls of the second trap was approximately 2.5 times higher than that in the first trap. The narrow cylindrical trap is depicted by dashed lines in Fig.5.

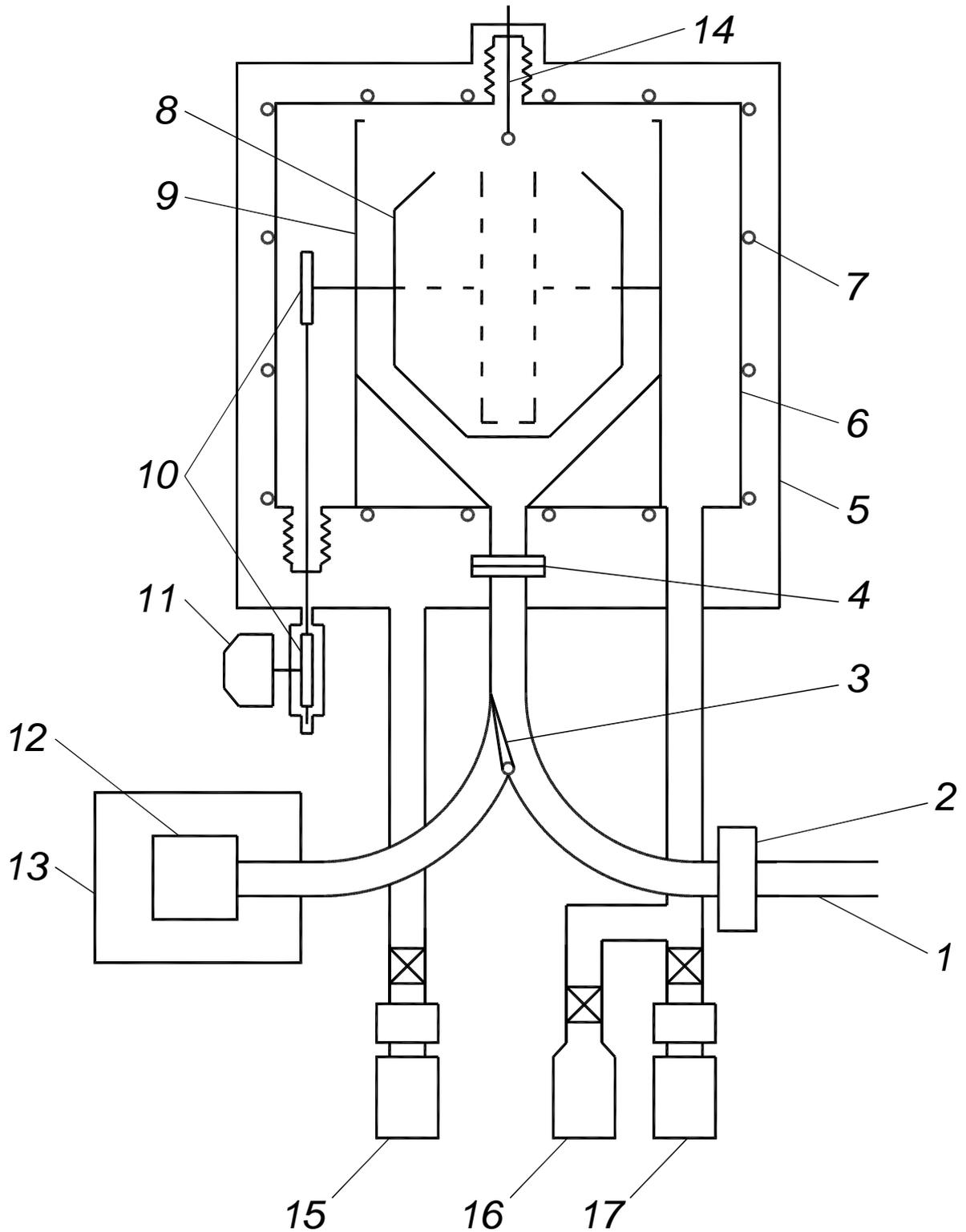

Fig.5. Schematic of the gravitational UCN storage system: 1, input neutron guide for UCNs; 2, inlet valve; 3, selector valve (shown in the position in which the trap is being filled with neutrons); 4, foil unit; 5, vacuum volume; 6, separate vacuum volume of the cryostat; 7, cooling system for the thermal shields; 8, UCN storage trap (the dashed lines depict a narrow cylindrical); 9, cryostat; 10, trap rotation drive; 11, step motor; 12, UCN detector; 13, detector shield; 14, vaporizer; 15, turbo-molecular pump with 80 K trap; 16, sorption pump; 17, turbo-molecular pump with 80 K trap.

The temperature dependence of UCN loss factor $\eta$ for different beryllium traps is shown in Fig.6. The curve 1 corresponds to beryllium-sputtered quasi-spherical trap without degassing.

This trap was made on copper. Besides all-metal beryllium trap was studied also (curve 3). Any questions about losses due to pinholes on the coated surface (assumed in work [1]) are excluded in these studies because of the copper critical energy higher than energy of stored neutrons. Moreover there are no questions for all-metall beryllium trap. After degasation and deuteration of the trap surface the curve 4 was obtained instead of curve 1. The temperature dependence of curve 4 corresponds to theoretical temperature dependence for beryllium, but there is additional temperature independent contribution $\sim 3 \cdot 10^{-5}$. It is small part but it is important. So, in work [2] problem of presence of hydrogen is solved by means of operation of deuteration.. The temperature-independent contribution of losses $3 \cdot 10^{-5}$ has been named by anomalous losses. Really, cross section in substance is at least 10 times less (see experimental curve 6).

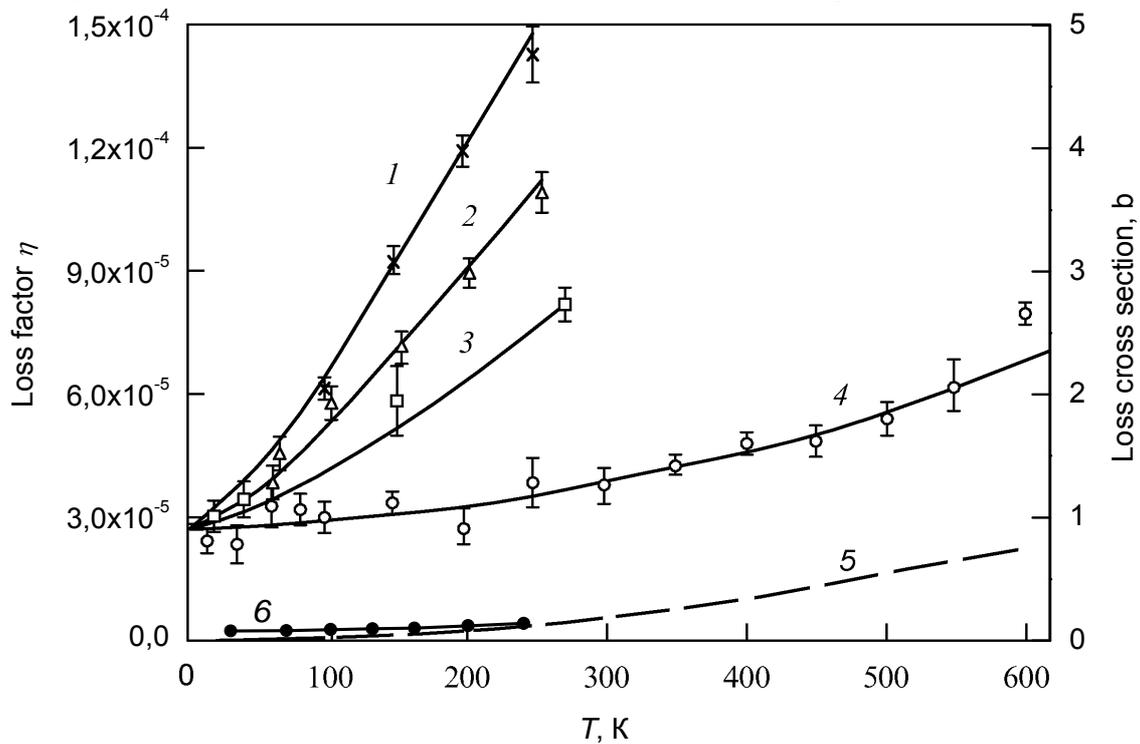

Fig. 6. Temperature dependence of the UCN loss factor $\eta$ for different beryllium traps: 1, beryllium-sputtered spherical trap without degassing; 2, degassed (5 hours at 250°C) beryllium-sputtered cylindrical trap; 3, degassed (8 hours at 300°C) all-metal beryllium trap; 4, degassed (28 hours at 350°C with a flow of He and $D_2$) beryllium-sputtered spherical trap; 5, theoretical temperature dependence calculated within the Debye model for beryllium; 6, experimental dependence of loss cross section in the neutron transmission experiment through the solid beryllium (this cross section includes elastic scattering on the inhomogeneity of the pressed beryllium).

Deuteration in work [1] has not given any result, and unexplained losses in work [1] in 27 times bigger, than anomalous losses in work [2]. Nevertheless, in work [1] the conclusion is drawn that in our work [2] anomalous losses have been connected with a film condensed on a surface at low temperatures since pumping was made by an oil diffusion pump. Actually pumping was made by means of turbo-molecular pump with a nitrogen trap. The idea of a hydrogen-containing film [1] on surface cannot be accepted for following reasons. First of all, the film ~100A at a room temperature can be only a film of substance with a low pressure of vapors otherwise it will be pumped away. Most possibly such film is an oil film. Such version does not correspond to vacuum pumping technology of works [2] and [1]. The idea of a film gives experimental dependence of losses $\mu \sim 1/\upsilon$, and for model of the dissolved hydrogen the dependence of losses is proportional $\upsilon$ at $\upsilon/\upsilon_c < 1$. Energy dependence of losses at a room temperature has been measured in work [2] by means of a gravitational spectrometer. It is presented in Fig.7 for a room temperature and for temperature 80K. In both cases of loss have dependence which have better fit when $\mu \sim \upsilon^{1.3}$ or $\mu \sim \upsilon^{1.5}$. This dependence strictly contradicts dependence $1/\upsilon$ for a film.

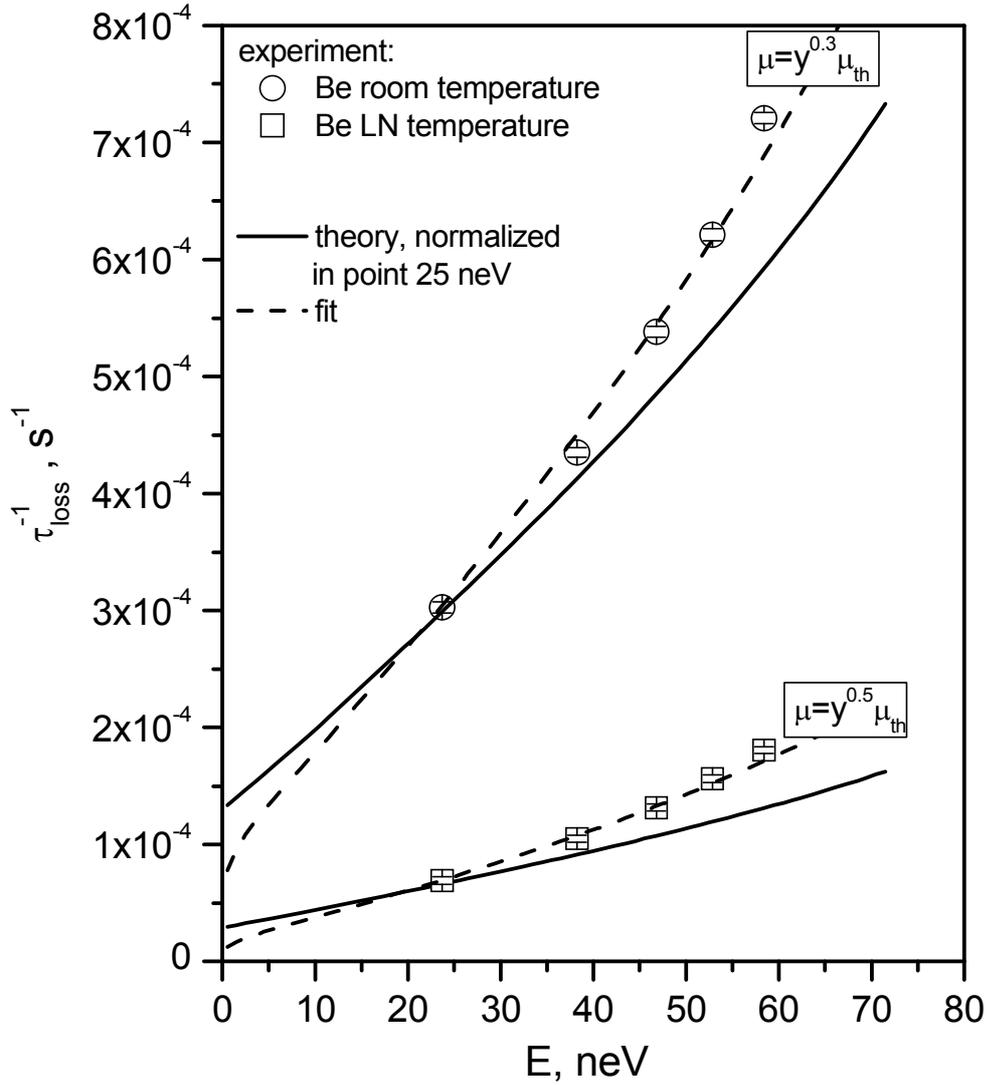

Fig. 7. Energy dependence of UCN loss rate for narrow Cu trap coated by beryllium.

As a whole it is necessary to conclude that the model of hydrogen-contained films on the substance surface, declared in work [1], has no relation to reality. The model of the dissolved hydrogen which is presented in Fig. 3 in work [1] is most probable. Thus, the experimental dependence of losses measured in work [1] more probably concerns slit deformation at cooling of traps instead of to temperature dependence of UCN losses on hydrogen.

The problem of anomalous losses on beryllium still demands the explanation. Though it is already clear that anomalous losses have no universal character since on the frozen low-temperature fomblin (LTF) the factor of losses is $2 \cdot 10^{-6}$ [4], i.e. almost 10 times less than on beryllium. These losses correspond to neutron inelastic scattering in LTF at temperature 100K. Our studies of the structure of beryllium show that there are defects of substance that can be observed by means of neutron transmission experiments. These studies have been carried out for different beryllium samples: pressed beryllium, quasi-single-crystal beryllium a melted beryllium using the passage of neutrons with velocity of 10-12 m/s. In this case above-barrier neutrons ($E_n > E_{Be}$) was used. The studies of beryllium coating were carried out with sub-barrier neutrons ($E_n < E_{Be}$) and Al or Si substrates [5]. It was shown that there are defects of material which can play considerable role in UCN interaction with surface. In this connection the idea of UCN localization around of defects has been proposed [6]. But here is very difficult question about the transition of UCN from vacuum state to the localized state in substance. Therefore the explanation of anomalous was not successful. Recently the work [7] was published where effect of UCN localization is explaining by resonance sub-barrier capture of UCN on defects of substance. This model is represented plausible.